# Skin-inspired in-sensor encoding of strain vector using tunable quantum geometry

*Zenglin Liu, Jingwen Shi, Jin Cao, Zecheng Ma, Zaizheng Yang, Yanwei Cui, Lizheng Wang, Yudi Dai, Moyu Chen, Pengfei Wang, Yongqin Xie, Fanqiang Chen, Youguo Shi, Cong Xiao, Shengyuan A. Yang, Bin Cheng\*, Shi-Jun Liang\*, Feng Miao\**

Z. Liu, J. Shi, Z. Ma, Z. Yang, Y. Cui, L. Wang, Y. Dai, M. Chen, P. Wang, Y. Xie, F. Chen, S. -J. Liang, F. Miao
National Laboratory of Solid State Microstructures, Institute of Brain-Inspired Intelligence, School of Physics, Collaborative Innovation Center of Advanced Microstructures, Nanjing University, 210093, Nanjing, China.
E-mail: sjliang@nju.edu.cn; miao@nju.edu.cn

B. Cheng
Institute of Interdisciplinary Physical Sciences, School of Science, Nanjing University of Science and Technology, 210094, Nanjing, China.
E-mail: bincheng@njust.edu.cn

J. Cao, C. Xiao, S. A. Yang
Institute of Applied Physics and Materials Engineering, University of Macau, 519000, Macau SAR, China.

Y. Shi
Institute of Physics, Chinese Academy of Sciences, 100190, Beijing, China.

Keywords: quantum geometry, strain engineering, nonlinear Hall effect, in-sensor compting, high-dimensional encoding, bionic tactile perception

**Abstract**

Human skin provides crucial tactile feedback, allowing us to skillfully perceive various objects by sensing and encoding complex deformations through multiple parameters in each tactile receptor. However, replicating this high-dimensional tactile perception with conventional materials' electronic properties remains a daunting challenge. Here, we present a skin-inspired method to encode strain vectors directly within a sensor. This is achieved by leveraging the strain-tunable quantum properties of electronic bands in the van der Waals topological semimetal $T_d$-WTe$_2$. We observe robust and independent responses from the second-order and third-order nonlinear Hall signals in $T_d$-WTe$_2$ when subjected to variations in both the magnitude and direction of strain. Through rigorous temperature-dependent measurements and scaling law analysis, we establish that these strain responses primarily stem from quantum geometry-related phenomena, including the Berry curvature and Berry-connection polarizability tensor. Furthermore, our study demonstrates that the strain-dependent nonlinear Hall signals can efficiently encode high-dimensional strain information using a single device. This capability enables accurate and comprehensive sensing of complex strain patterns in the embossed character "NJU". Our findings highlight the promising application of topological quantum materials in advancing next-generation, bio-inspired flexible electronics.

## 1. Introduction

Human skin, with its intricate network of tactile receptors, serves as a highly efficient system for tactile perception, allowing human beings to perceive and interact with the surrounding environment.[1] When the skin interacts with shaped objects, tactile receptors encode high-dimensional information into bioelectrical signals that carry multiple parameters of skin deformation,[2,3] as illustrated in **Figure 1a**. This encoding scheme forms the basis for biological systems to perform complex tactile perception tasks, such as recognizing the surface textures and shapes of objects.[4-6] Utilizing the physical properties of intelligent electronic materials to emulate tactile perception of biological systems could revolutionize the way artificial systems sense and monitor surrounding environment,[7-10] which is expected to play a key role in practical applications such as augmented reality,[11,12] medical equipment,[13-15] and biomimetic robot.[16] However, it remains a grand challenge to achieve biomimetic tactile perception with human-level dexterity.[17] This mainly stems from the limitations of conventional strain-sensing materials,[18-23] which can only respond to either the magnitude or direction of the strain vector and cannot emulate human skin's ability to sense deformation information involving multiple parameters through encoding. Exploiting novel physics phenomena of quantum materials that can couple multiple degrees of freedom may provide a promising approach to mimic the encoding schemes of tactile receptors. Recent advancements in the electronic band engineering of two-dimensional quantum materials demonstrate the potential of utilizing tunable electronic band properties as highly sensitive probes for external field detection.[24-36] Among these properties, quantum geometry, which describes the geometric properties of electronic wave function,[37-39] can be significantly modulated by applied electric or magnetic fields.[40-48] Such unique properties exhibit great potential for simultaneous sensing of multiple external field parameters, such as high-dimensional optical information.[36] Beyond optical information of multiple degrees of freedom, exploration of strain-tunable quantum geometry, as alternative degrees of freedom to encode high-dimensional strain information, allows for the achievement of electronic skins with functionality beyond that based on the conventional electronic materials.

In this work, we introduce a novel approach inspired by human skin to encode strain vectors directly within a sensor. Leveraging the unique quantum geometric properties of few-layer samples of the van der Waals topological semimetal $T_d$-WTe$_2$, we develop a device that utilizes second-order and third-order nonlinear Hall voltage signals as independent parameters for encoding high-dimensional strain information. By modulating both the magnitude and direction of strain, we show that these signals accurately reflect the applied strain, enabling a single device to simultaneously sense and encode high-dimensional information in the embossed characters. Our findings underscore the tunable nature of quantum geometry in response to strain vectors, pointing towards exciting possibilities for replicating the intricate sensitivity of human tactile perception using quantum materials.

## 2. Results

### 2.1. Encoding of strain vector using strain-tunable quantum geometry

To demonstrate how strain-induced changes in quantum geometry can emulate the encoding mechanisms akin to tactile receptors, we utilize a 2D tilted Dirac cone model to calculate the distribution of the Berry curvature (BC) $\Omega_z$ and Berry-connection polarizability tensor (BPT) $\overleftrightarrow{G}$, with corresponding results shown in **Figure 1b,c**. These geometric quantities are prominent manifestations of the band geometric properties in quantum materials and play a crucial role in the nonlinear transport behaviors.[42-52] Moreover, their momentum space distribution heavily relies on the lattice structure,[53] making them highly responsive to the changes in lattice parameters caused by strain. As depicted in Figure 1b, we observe $\Omega_z$ concentrating around the central point, with significantly varied distributions across different strain vectors. Similarly, Figure 1c illustrates distinct patterns in the distribution of $G_{xy}$ (the component of BPT) for different strain vectors, which are uncoupled from the response of the $\Omega_z$ distribution under the same strain. These observations indicate that the two geometric quantities originating from quantum geometry can serve as independent degrees of freedom in response to the applied strain vectors. To further show that these two geometric quantities respond significantly and independently to different strain vectors, we present the distribution of their partial derivatives in Figure S1. Notably, such a strain response is unprecedented, as quantum geometry provides multiple degrees of freedom to couple multi-dimensional information involving both the magnitude and direction of strain vectors, and naturally encodes it in the momentum space distribution of these geometric quantities. This characteristic is akin to the biological encoding performed by a tactile receptor, therefore leveraging this strain-tunable quantum geometry enables the establishment of a new method for encoding high-dimensional strain vector information using a single device.

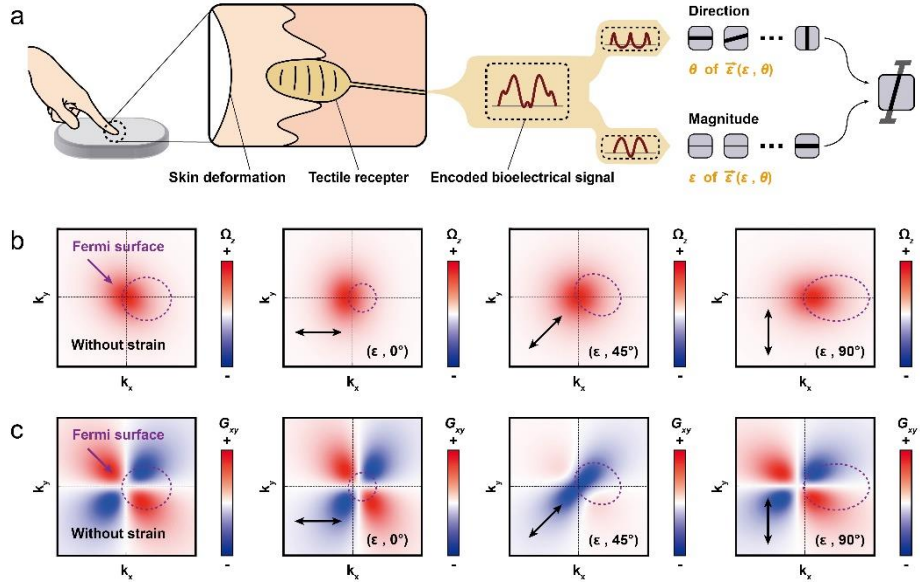

**Figure 1.** Encoding of strain vector using strain-tunable quantum geometry. a) Conceptual illustration of biological encoding scheme. Tactile receptors generate encoded bioelectrical signals conveying information about the direction and magnitude of skin deformation,

corresponding to the direction and magnitude of the strain vector. This information forms the foundation for the tactile perception of the italicized character "I". b) The distribution of the Berry curvature without strain is shown on the left, while the three images on the right depict the distributions of the Berry curvature when uniaxial strain vectors (black arrow) are applied at in 0°, 45° and 90°, respectively. c) The distribution of the Berry-connection polarizability tensor components $G_{xy}$ without strain is shown on the left, while the three images on the right depict the distributions of the Berry-connection polarizability tensor components $G_{xy}$ when uniaxial strain vectors (black arrow) are applied at 0°, 45°and 90°, respectively. The geometric quantities of electronic bands are calculated from the two-dimensional tilted Dirac model (see Experimental Section and Figure S1 for details), with the purple dashed line indicating the position of the Fermi surface.

**2.2. Characterization of strain-tunable quantum geometry by nonlinear Hall signals**

We experimentally investigate the strain response of quantum geometry in the topological semimetal $T_d$-WTe$_2$ using second-order and third-order nonlinear Hall (NLH) effects as probes. The distribution of BC and BPT can be modulated by the magnitude and direction of strain vectors as mentioned above, thereby facilitating the strain information encoding through quantum geometry. Such strain tuning of quantum geometry can be monitored by measuring the second-order and third-order terms of NLH effects, which can be illustrated by $j_y^{2\omega} = \chi_{yxx}^{2\omega} E_x E_x$ and $j_y^{3\omega} = \chi_{yxxx}^{3\omega} E_x E_x E_x$, respectively. Here, $\chi_{yxx}^{2\omega}$ and $\chi_{yxxx}^{3\omega}$ represent the component of the second-order and third-order nonlinear Hall conductivity tensors, and their relationships with BC and BPT can be expressed as $\chi_{yxx}^{2\omega} = -(\tau/2)\int_k f_0 \partial_x \Omega_z$ and $\chi_{yxxx}^{3\omega} = -\tau \int_k f_0 \partial_x^2 G_{xy} + (\tau/2) \int_k \partial_\varepsilon^2 f_0 v_x v_y G_{xx}$. To characterize the strain response of second-order and third-order NLH signals, we fabricate the few-layer $T_d$-WTe$_2$ device in a Hall bar configuration on the flexible polyimide (PI) substrate, as shown in **Figure 2a**. The optical image and height profile of a typical $T_d$-WTe$_2$ device are shown in Figure S2. We confirm that the channel direction of the $T_d$-WTe$_2$ device is along the *a*-axis using the polarized Raman spectroscopy presented in Figure S3, and define it as the 0° direction of the strain vector. More details about the device fabrication and configuration are provided in Experimental Section. We then conduct standard nonlinear transport measurements on $T_d$-WTe$_2$ devices, maintaining the harmonic driving current *I* along *a*-axis to simultaneously monitor the longitudinal voltage signal at fundamental frequency $V_{xx}^{\omega}$, and the transverse voltage signals at second-harmonic and third-harmonic frequencies (denoted as $V_{xy}^{2\omega}$ and $V_{xy}^{3\omega}$). Meanwhile, the homemade strain setup allows us to apply strain to the devices, adjusting the magnitude and direction of strain vectors by bending and rotating the PI substrate (Figure S4). As a result, we obtain the $V_{xy}^{2\omega}$ versus $I^2$ and $V_{xy}^{3\omega}$ versus $I^3$ curves under different strain vectors at 0° (*a*-axis), 45°, 90° (*b*-axis) with different magnitudes, which are depicted in **Figure 2b–g**. All these curves exhibit linear behavior, corresponding to second-order and third-order nonlinear charge responses, respectively.

To precisely characterize the strength of NLH effect, we extract intensive quantities $E_{xy}^{2\omega}/E_{xx}^2$ and $E_{xy}^{3\omega}/E_{xx}^3$,[44,45] so that the impact of longitudinal resistivity changes on strain response in the NLH signals can be eliminated. Here, $E_{xx}$ is the longitudinal electric field defined by $V_{xx}^{\omega}/L$, while $E_{xy}^{2\omega} = V_{xy}^{2\omega}/W$ and $E_{xy}^{3\omega} = V_{xy}^{3\omega}/W$ denote the transverse electric fields of second-harmonic and third-harmonic frequencies, respectively, with *L* and *W* the channel length and width. By fitting the slope of the $E_{xy}^{2\omega}$ versus $E_{xx}^2$ and $E_{xy}^{3\omega}$ versus $E_{xx}^3$ curves (Figure S5), we obtain the $E_{xy}^{2\omega}/E_{xx}^2$ and $E_{xy}^{3\omega}/E_{xx}^3$ under different strain vectors, as depicted in **Figure 2h,i**. We observe the uncoupled behavior modulated by strain vectors on the second-order and third-order NLH signals. To be specific, the second-order NLH signal $E_{xy}^{2\omega}/E_{xx}^2$ increases with strain vectors applied at 45° and 90°, while the third-order NLH signal $E_{xy}^{3\omega}/E_{xx}^3$ exhibits pronounced decreases when strain vectors are applied at 0° and 45°. This observation indicates that the responses of the second-order and third-order NLH signals to either magnitude or direction of the strain vector are independent, moreover such strain responses are robust and reproducible across several devices (Figure S6). It is noteworthy that NLH signals with multiple orders can reveal multiple independent degrees of freedom provided by quantum geometry, thus we can characterize strain vector information encoded by quantum geometry through the second-order and third-order NLH signals.

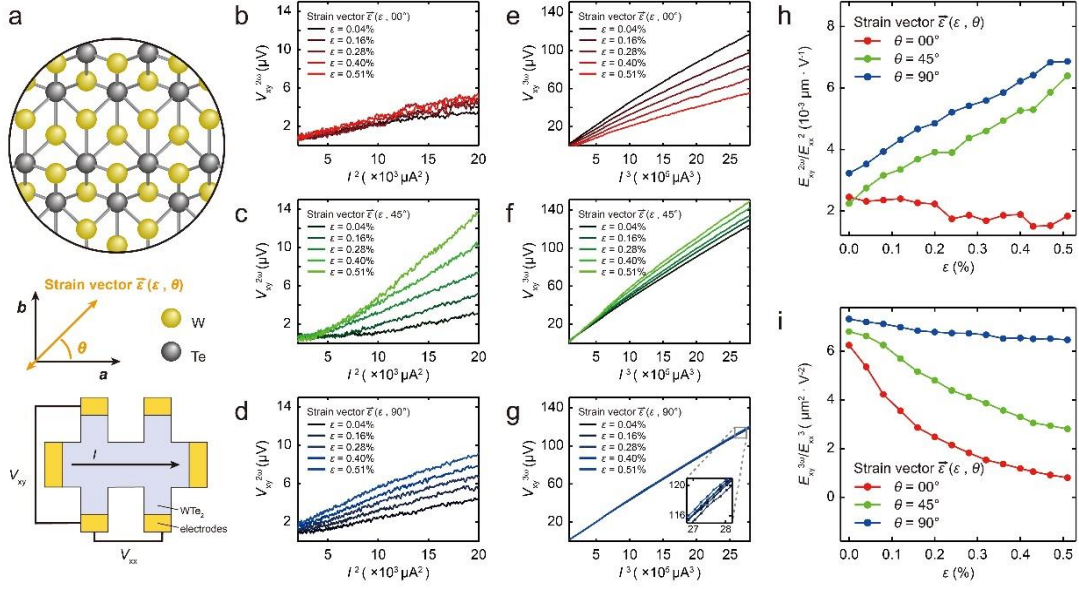

**Figure 2.** Characterization of strain-tunable quantum geometry by nonlinear Hall signals. a) Top view of the atomic lattice of $T_d$-WTe$_2$ and schematic of the $T_d$-WTe$_2$ device for electric transport measurements. The black arrows represent directions along *a*-axis (the direction of the applied current) and *b*-axis respectively, while the orange arrows depict the direction of the in-plane uniaxial strain vector. b,c,d) Second-harmonic voltage signal $V_{xy}^{2\omega}$ depends linearly on the square of harmonic current amplitude *I* when different magnitudes of strain vectors are applied at 0° (b), 45° (c) and 90° (d). e,f,g) Third-harmonic voltage signal $V_{xy}^{3\omega}$ depends linearly on the cube of harmonic current amplitude *I* when different magnitudes of strain vectors are applied at 0° (e), 45° (f) and 90° (g). The insert plot in (g) is enlarged from the gray rectangular area. h) The second-order nonlinear Hall signals $E_{xy}^{2\omega}/E_{xx}^2$ are plotted as functions of the magnitude of strain vector applied in different directions. i) The third-order nonlinear Hall signals $E_{xy}^{3\omega}/E_{xx}^3$ are plotted as functions of the magnitude of strain vector applied in different directions.

### 2.3. Encoding mechanism revealed by temperature dependence measurement

To investigate the dominant mechanism for performing the encoding of strain vectors, we conduct the temperature dependence measurements of NLH signals. We fix the strain at 0% and 0.51% (the directions include 0°, 45°, and 90°) and record the variation of the second-harmonic (third-harmonic) transverse electric fields $E_{xy}^{2\omega}$ ($E_{xy}^{3\omega}$) with the square (cube) of longitudinal electric field at different temperatures, with corresponding results shown in Figure S7. The $E_{xy}^{2\omega}$ versus $E_{xx}^2$ and $E_{xy}^{3\omega}$ versus $E_{xx}^3$ curves obtained from temperature dependence measurements exhibit linear behavior, allowing us to extract the second-order NLH signal $E_{xy}^{2\omega}/E_{xx}^2$ as well as the third-order NLH signal $E_{xy}^{3\omega}/E_{xx}^3$ by fitting the slopes of the curves. **Figure 3a,b** respectively illustrate the temperature dependence of the extracted NLH signals ($E_{xy}^{2\omega}/E_{xx}^2$ and $E_{xy}^{3\omega}/E_{xx}^3$) under strain modulation of 0% and 0.51%, with directions including 0°, 45°, and 90°. We note that both the second-order and third-order NLH signals decrease as temperature increases, with the third-order NLH signal $E_{xy}^{3\omega}/E_{xx}^3$ decreasing more rapidly. Despite the reduction in strain responses of second-order and third-order NLH signals with temperature changes consistent with the temperature dependence of NLH signals, their independent characteristic remains preserved. Similar results have also been observed in the strain response measurements with the temperature range up to 260 K (Figure S8), indicating that the mechanism generating the strain response remains dominant over the measured temperature range.

We then characterize the scaling laws $E_{xy}^{2\omega}/E_{xx}^2 = \xi^{2\omega}\sigma^2 + \eta^{2\omega}$ and $E_{xy}^{3\omega}/E_{xx}^3 = \xi^{3\omega}\sigma^2 + \eta^{3\omega}$ to analyze the results of temperature-dependent measurements, where the longitudinal conductivity $\sigma$ is obtained from $\sigma = I(L/Wd)/V_{xx}^\omega$ with $d$ the sample thickness.[44,45,54] The coefficients $\xi^{2\omega}$ and $\xi^{3\omega}$ involve the mixing contributions from various skew scattering processes,[46,55] which usually play crucial roles in disorder electronic systems. Meanwhile, the coefficients $\eta^{2\omega}$ and $\eta^{3\omega}$ are mainly contributed by the distribution of geometric quantities BC and BPT, representing the intrinsic mechanisms. The experimental data of $E_{xy}^{2\omega}/E_{xx}^2$ and $E_{xy}^{3\omega}/E_{xx}^3$ with respect to $\sigma^2$ for different strain vectors are illustrated in **Figure 3c,d**, respectively, showing that all curves fit well with the scaling laws. We obtain the coefficients $\xi^{n\omega}$ and $\eta^{n\omega}$ related to scattering mechanisms and intrinsic mechanisms by fitting the curves modulated by different strain vectors, with the results listed in Table S1. Notably, in the absence of strain, the coefficients $\eta^{n\omega}$ contributed by the intrinsic mechanism of the second-order and third-order NLH effects are $\eta^{2\omega} = 6.554\times10^{-4}$ μm·V$^{-1}$ and $\eta^{3\omega} = -8.653$ μm$^2$·V$^{-2}$. These values are close to the coefficients reported previously in topological semimetals,[44,45] consistent with the widely acknowledged scenario that significant BC and BPT contribute to the NLH signals in these materials. Further, we observe that under the modulation of strain vectors, the changes in the coefficients $\eta^{2\omega}$ and $\eta^{3\omega}$ are more pronounced compared to those in the coefficients $\xi^{2\omega}$ and $\xi^{3\omega}$. Remarkably, the coefficients $\xi^{2\omega}$ and $\xi^{3\omega}$ exhibit similar responses with respect to the direction of strain vector, whereas the coefficients $\eta^{2\omega}$ and $\eta^{3\omega}$ show distinct response as we change the strain direction. Such independence of the strain response of $\eta^{2\omega}$ and $\eta^{3\omega}$ is consistent with the strain direction dependent responses of $E_{xy}^{2\omega}/E_{xx}^2$ and $E_{xy}^{3\omega}/E_{xx}^3$ already demonstrated in Figure 2h,i, indicating that the strain responses of the second-order and third-order NLH signals primarily stem from the variations in the distribution of BC and BPT modulated by strain, respectively. These observations support the characterization of quantum geometry through NLH signals, thereby enabling us to encode high-dimensional information of strain vectors using strain-tunable quantum geometry.

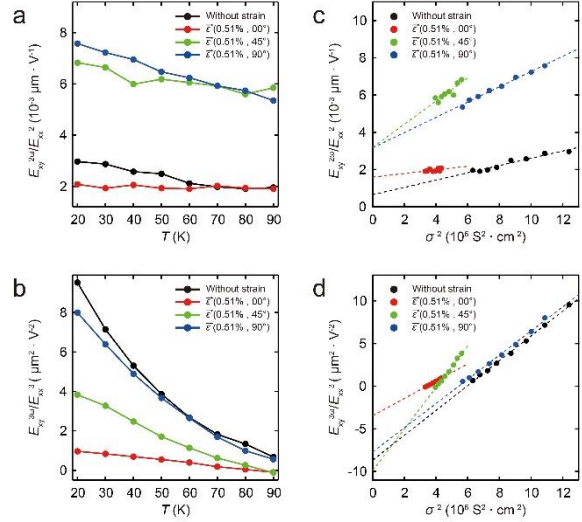

**Figure 3.** Encoding mechanism revealed by temperature dependence measurement. a) The second-order nonlinear Hall signals $E_{xy}^{2\omega}/E_{xx}^2$ are plotted as a function of temperature with strain vectors applied in different directions or without strain. b) The third-order nonlinear Hall signals $E_{xy}^{3\omega}/E_{xx}^3$ are plotted as a function of temperature with strain vectors applied in different directions or without strain. c) The second-order nonlinear Hall signals $E_{xy}^{2\omega}/E_{xx}^2$ are plotted as a function of the square of the longitudinal conductivity with strain vectors applied in different directions or without strain. d) The third-order nonlinear Hall signals $E_{xy}^{3\omega}/E_{xx}^3$ are plotted as a function of the square of the longitudinal conductivity with strain vectors applied in different directions or without strain. The dashed lines in (c) and (d) are linear fits to the experimental data (dots). The coefficients extracted from these fitted curves are presented in Table S1.

### 2.4. Proof-of-concept demonstration of quantum geometry encoder

Finally, we explore the implementation of a skin-inspired encoding scheme for high-dimensional sensing and in-sensor encoding of strain vectors. In our scheme, a single $WTe_2$ device serves as a quantum geometry encoder to emulate the tactile perception process of tactile receptors in human skin, transmitting information about the magnitude and direction of strain vectors through encoded electrical signals converted by NLH effects. To provide a proof-of-concept demonstration, we present the encoding of high-dimensional information in the embossed characters "NJU", as illustrated in **Figure 4a**. Upon contact with the embossed characters, our device undergoes deformation stretching along a specific direction and generates a strain response that corresponds to such deformation in the Hall signal. Similar to the encoded bioelectrical signals generated by tactile receptors, the second-harmonic and third-harmonic components of the Hall signal can implicitly contain information about the magnitude and direction of the strain vector encoded by quantum geometry. To visually display the high-dimensional information of strain vectors carried by the encoded Hall signal, we employ an artificial neural network (ANN) which has been widely used as a tool for information extraction tasks. The inputs to the ANN consist of the second-harmonic or third-harmonic voltage signal derived from the Hall signal. After processing by the ANN, we examine the information regarding the magnitude and direction of the strain vector obtained from the ANN's output, thereby evaluating the capability of quantum geometry encoder to emulate tactile receptors in human skin for sensing and encoding high-dimensional information. More details about the construction of the ANN and the implementation of the quantum geometry encoder are provided in Experimental Section.

To further illustrate how the quantum geometry encoder achieves high-dimensional encoding of strain vectors, we provide a comprehensive description for the generation and evolution of encoded signals within in the 4×4 pixels detected region, including a part of the character "U" and the block's edge, as shown in **Figure 4b**. We employ a single device to traverse the detected region, which encompasses different strain directions (0°, 45°, 90°) and magnitudes (0.2%, 0.4%, 0.5%). In this process, we fix the driving current (140 μA) in the device and record outputs voltage signals containing the fundamental, second-harmonic, and third-harmonic frequencies. Notably, the fundamental longitudinal voltage signal $V_{xx}^{\omega}$ induced by the piezoresistive effect is only sensitive to strain vectors close to 90° within the detection region. This observation indicates that $V_{xx}^{\omega}$ is only effective for detecting strain in the specific direction aligned with the device's channel, indicating that the piezoresistive effect is limited in realizing high-dimensional sensing. In contrast, the responses of the second-harmonic and third-harmonic transverse voltage signals ($V_{xy}^{2\omega}$ and $V_{xy}^{3\omega}$) to the strain vectors are not required to be confined to any specific direction. Moreover, the strain responses of $V_{xy}^{2\omega}$ and $V_{xy}^{3\omega}$ are independent of each other, thus they can serve as multiple independent degrees of freedom to contain multi-dimensional strain vector information.

We next train the ANN using the strain responses of $V_{xy}^{2\omega}$ and $V_{xy}^{3\omega}$ (Figure S9) as the dataset, to extract information about magnitude and direction of strain vectors. As shown in **Figure 4c**, we investigate the performance of the trained ANN using 21 representative datasets. Additionally, the accuracy and mean square error for the full dataset related to the training of ANN are presented in Figure S10. By processing with the trained ANN, the NLH voltage signal at each pixel of detected region is converted into normalized magnitude and direction of the strain vector. These outputs align with the distribution of strain vectors in the 4×4 pixels detected region, confirming the capability of high-dimensional encoding based on the strain-tuning of quantum geometry properties in the WTe$_2$ device. We further show the encoding results for the embossed characters "NJU" in **Figure 4d,e**, where the signal processing (as shown in Figure S11) is similar to that in the 4×4 pixels detected region. Our scheme thus achieves the encoding of high-dimensional information about the strain vector through a single device, enabling comprehensive sensing of both magnitude and direction.

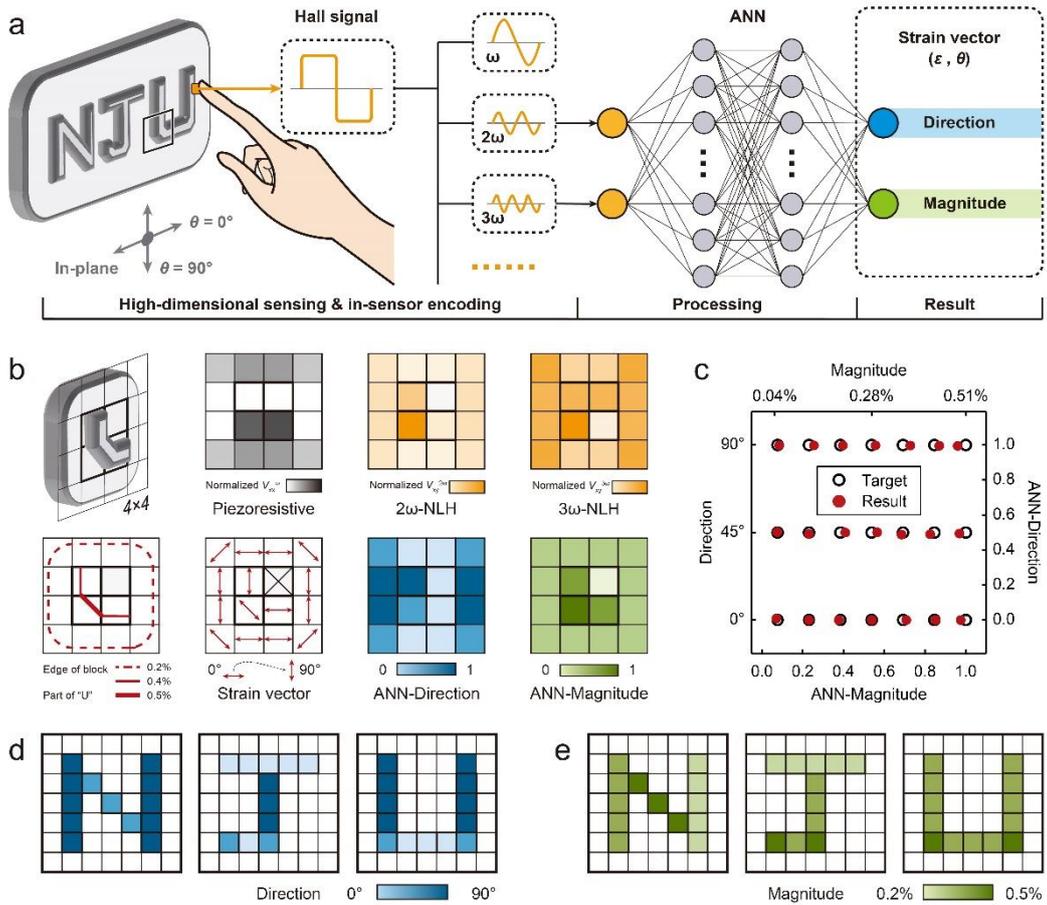

**Figure 4.** Proof-of-concept demonstration of quantum geometry encoder. a) Conceptual illustration of the skin-inspired encoding scheme. The second-harmonic and third-harmonic

components of the Hall signal generated by the quantum geometry encoder, implicitly encode the direction and magnitude of strain vectors. These harmonic signals are then processed by a trained artificial neural network comprising two input neurons, two hidden layers, and two output neurons. The input of ANN incorporates the measured $V_{xy}^{2\omega}$ and $V_{xy}^{3\omega}$, while the output of ANN provides the corresponding direction and magnitude of strain vector. The gray coordinate axis in the illustration represents the direction of the in-plane strain vector. b) The 4×4 pixels detected region is enlarged from the black rectangular area in (a), containing the part of character "U" and the block's edge. The mappings of normalized voltage signal at fundamental, second-harmonic, and third-harmonic frequencies are displayed in the three graphs on the right side of the upper panel. In the lower panel, the distribution of strain vectors for this region is shown on the left side and the corresponding direction and magnitude of strain vector output by ANN are shown on the right side. c) Strain vectors output by the ANN for verification (red dots) and their corresponding training targets (black circles). d,e) Simulation results for encoding the strain vector in embossed characters "NJU" are provided through the mappings of the direction (d) and the magnitude (e).

## 3. Conclusion

In summary, we propose and implement a skin-inspired in-sensor encoding scheme by exploring the exceptional strain tunability of quantum geometry. Our scheme leverages the quantum geometry of electronic bands in the van der Waals topological semimetal $T_d$-WTe$_2$, enabling a single device to perform high-dimensional sensing and in-sensor encoding of strain vectors. This capability not only allows for the comprehensive recording of multi-parametric tactile information but also reduces information complexity through natural encoding, which is inaccessible with conventional strain sensing. Our work spotlights the immense potential of exploiting the unique properties of quantum materials to construct biomimetic perception with human-level dexterity.

## 4. Experimental Section

**Device fabrication**

Considering the sensitive chemical nature of the few-layer $T_d$-WTe$_2$, the device fabrication process was completed in an inert gas environment within a glovebox. Thin flakes of WTe$_2$ were mechanically exfoliated from the bulk crystal onto O$_2$ plasma-cleaned 300-nm SiO$_2$/Si wafers. Optical contrast was utilized to identify a suitable thin flake of WTe$_2$, and the crystalline direction of the identified WTe$_2$ was characterized using WITec optical systems. After that we transfer the identified WTe$_2$ onto the polyimide (PI) substrate with prepatterned bottom electrodes (Ti 2 nm/Au 20 nm). In this step, the *a*-axis of WTe$_2$ was aligned with the channel direction of the bottom electrodes. Standard electron beam lithography and e-beam evaporation technique were utilized to fabricate the top electrode (Ti 5 nm/Au 55 nm) for anchoring WTe$_2$ to the PI substrate. Additionally, a thick layer of h-BN was stacked onto WTe$_2$ for protection. Poly propylene carbonate (PPC) and polydimethylsiloxane (PDMS) were used to pick up and release WTe$_2$ and h-BN flakes. The thickness of WTe$_2$ was characterized using a Bruker Multimode Atomic Force Microscope.

**Electrical measurements**

The homemade strain setup was loaded in an Oxford TeslatronPT system, which provides sample temperatures between 2 K and 300 K. A harmonic current with 137.77 Hz as fundamental frequency was applied to the devices. The longitudinal and transverse voltage signals at frequencies from fundamental to third-harmonic were measured using lock-in amplifiers (SR830).

**2D tilted Dirac cone model**

Using a two-dimensional (2D) Dirac model, we illustrate the tunability of the intrinsic band quantities under external strain. We investigate the Berry curvature dipole[49,50]

$$D_\alpha(\boldsymbol{k}) = \partial_\alpha \Omega_z, \tag{1}$$

where $\Omega_z$ is the Berry curvature (BC), and the Berry-connection polarizability tensor (BPT)[51,52]

$$G_{ab}(\boldsymbol{k}) = 2\mathrm{Re} \sum_{n \neq 0} \frac{(\mathcal{A}_a)_{0n} (\mathcal{A}_b)_{n0}}{\varepsilon_0 - \varepsilon_n}, \tag{2}$$

where $(\mathcal{A}_a)_{mn} = i\langle u_m | \partial_a u_n \rangle$ is the interband Berry connection, and $\varepsilon_n$ is the band energy. They are related to the time-reversal (TR) invariant second-order and third-order Hall effect,[43-45] respectively. The 2D Dirac model near $\boldsymbol{Q}$ point is written as

$$H(\boldsymbol{q}) = \omega q_x + v(q_x \sigma_x + q_y \sigma_y) + \Delta \sigma_z, \tag{3}$$

where the momentum $\boldsymbol{q} = \boldsymbol{k} - \boldsymbol{Q}$. We use a tilt parameter $\omega = 0.05$ eV, Fermi velocity $v = 0.1$ eV·Å, and $\Delta = 0.1$ eV. The model has $\mathcal{TM}_x$ symmetry. Under the in-plane strain

$$\boldsymbol{\mathcal{E}}(\theta) = R_\theta \begin{pmatrix} \varepsilon & 0 \\ 0 & -\nu\varepsilon \end{pmatrix} R_\theta^{-1}, \tag{4}$$

the reciprocal space deforms as

$$k' = (1 - \boldsymbol{\mathcal{E}}^T)k. \tag{5}$$

Here, $v = 0.165$ is the Poisson ratio, and $R_\theta$ is rotation matrix. Also, the strain will induce a gauge field $\boldsymbol{A} = \frac{\beta}{a}(\varepsilon_{xx} - \varepsilon_{yy}, -2\varepsilon_{xy})$ due to the couplings of long-wavelength phonons to electrons,[56,57] where we take $\beta/a = 1.1$. Then, the Hamiltonian under strain can be derived as

$$H'(\boldsymbol{q}) = H\left[(1+\mathcal{E})q + \boldsymbol{A} - \mathcal{E}\boldsymbol{Q}\right]. \tag{6}$$

The calculated distributions of BC and BPT for the valence band with applied in-plane strain are shown in Figure 1b,c and Figure S1. The Fermi surfaces are plotted in purple line. To break the mirror symmetry, the strain should be applied in a direction that is neither parallel nor perpendicular to the mirror line. For example, the second-order Hall conductivity

$$\chi^{(2)}_{xyy} = -\frac{\tau}{2}\int\frac{d\boldsymbol{k}}{(2\pi)^2} f_0 \partial_y \Omega_z, \tag{7}$$

is obtained by integrating out $D_y$ of the occupied states. Here, $\tau$ is the relaxation time, and $f_0$ is the Fermi-Dirac function. Such a tensor element is responsible for the second-order Hall effect for applied electronic field along $y$ direction, and is forbidden by $\mathcal{TM}_x$ symmetry in the Dirac model, as shown in Figure S1b. In the presence of strain along $x$ or $y$ directions, one can identify the deformation of the distribution of $D_y$ as well as the Fermi surface, however, the mirror symmetry is preserved, resulting in vanished $\chi^{(2)}_{xyy}$. The distribution is asymmetric for $\theta = \pi/4$, leading to nonzero $\chi^{(2)}_{xyy}$. The strain induced nonzero $\chi^{(2)}_{xyy}$ follows a $\pi$-period pattern with respect to $\theta$. It is also the case for the third-order Hall conductivity

$$\chi^{(3)}_{yxxx} = -\tau\int\frac{d\boldsymbol{k}}{(2\pi)^2} f_0 \partial_x^2 G_{xy} + \frac{\tau}{2}\int\frac{d\boldsymbol{k}}{(2\pi)^2} \partial_\varepsilon^2 f_0 v_x v_y G_{xx}, \tag{8}$$

where $v_x$ or $v_y$ is the group velocity. It should be noted despite our use of a 2D Dirac model lacking TR symmetry, the TR partner of the Dirac cone will contribute the same to the TR-invariant nonlinear Hall conductivity.

**Determination of strain vector**

The WTe$_2$ device on the polyimide (PI) substrate was mounted onto a homemade strain setup featuring a three-point bending configuration. By pushing the horizontal round rod positioned along the midline of the back of PI substrate, the setup induced bending of the PI substrate, uniformly stretching its top surface, as depicted in Figure S4a. This action generated a uniaxial tensile strain vector perpendicular to the round rod, effectively transferring strain to the WTe$_2$. The piezoelectric displacer is responsible for actuating the round rod, while its sensor accurately measures the vertical height $h$ of the PI substrate being lifted.[58] The side view of the curved PI substrate is depicted in Figure S4e, where $t = 0.2$ mm is the thickness of the PI substrate, $r$ is the radius of curvature, and $w = 5$ mm is half the distance between the two fixed ends (the green block in Figure S4a). Based on definition of strain vector and the geometric relationship in our strain setup, the magnitude of strain $\varepsilon$ could be written as

$$\varepsilon = \frac{l - l_0}{l_0} = \frac{r2\alpha - \left(r - \frac{t}{2}\right)2\alpha}{\left(r - \frac{t}{2}\right)2\alpha} = \frac{t}{2r - t}, \tag{9}$$

where $l$ is the length of the stretched upper surface, $l_0$ is the original length and $\alpha$ is half of the angle of the arc. Because of the geometric relationship $r^2 = w^2 + (r - h)^2$, we can obtain

$$\varepsilon = \frac{t}{h + \frac{w^2}{h} - t}, \tag{10}$$

where $t$ and $w$ are fixed values. Such relationship between the magnitude of strain $\varepsilon$ and the vertical height $h$ is illustrated in Figure S4f, demonstrating an approximately linear correlation. The measurements of four-probe resistance under applied and removed strain at low temperature are depicted in Figure S4g, providing the validity of estimation and the robustness of our homemade strain setup.

**Construction of ANN and implementation of strain vector encoding**

The artificial neural network (ANN) employed in this work comprises an input layer (2×1), two fully connected layers (128×1 and 64×1) and an output layer (2×1). The weight parameters were optimized in the training process through back propagation and the loss function (Mean square error, MSE) was minimized by the gradient descent. The sigmoid activation function was used for two hidden layers and output layer to generate the outputs. A total of 39 original nonlinear Hall voltage signals and their corresponding strain vectors were measured for training and testing, as illustrated in Figure S9. The dataset includes 4290 nonlinear Hall voltage signals, which generated by cubic interpolation from the original 39 measured results, and they were randomly divided into training set (91%) and testing set (9%). Additionally, the validation dataset comprising 21 experimental data was constructed to verify the credibility. In the training process, the second-harmonic voltage signal $V_{xy}^{2\omega}$ and the third-harmonic voltage signal $V_{xy}^{3\omega}$ from each nonlinear Hall voltage signal are input into the ANN, generating corresponding outputs ($\varepsilon$, $\theta$). As the number of epochs increases the neural network was trained to obtain the optimal weight configuration, as illustrated in Figure S10. To demonstrate the performance of strain vector encoding, we generated three embossed letters "N", "J" and "U", whose topography contains five different combinations of strain information ((0.20%, 0°), (0.20%, 90°), (0.40%, 0°), (0.40%, 90°) and (0.51%, 45°)). Each letter was mapped to the 7×7 array of pixels, as illustrated in Figure S11. After loading the strain signals representing the embossed letters pixel-by-pixel, the strain parameters were simultaneously perceived and encoded by the $WTe_2$ device. Subsequently, the collected nonlinear Hall signals from the device were introduced to the ANN to extract the magnitude and direction of strain vector. To substantiate the performance of strain vector encoding, the simulation results for encoding strain information are provided in Figure S4d,f. All the algorithms were written with Python on the PyCharm platform.

## Supporting Information

Supporting Information is available from the Wiley Online Library or from the author.

## Acknowledgements

Z. Liu and J. Shi contributed equally to this work. This work was supported in part by the National Key R&D Program of China under Grant 2023YFF1203600, the National Natural Science Foundation of China (62122036,12322407, 62034004, 61921005, 12074176), the National Key R&D Program of China under Grant 2023YFF0718400, the Leading-edge Technology Program of Jiangsu Natural Science Foundation (BK20232004), the Strategic Priority Research Program of the Chinese Academy of Sciences (XDB44000000), Supported by the Fundamental Research Funds for the Central Universities (02042103031). F. Miao and S. -J. Liang would like to acknowledge support from the AIQ Foundation and the e-Science Center of Collaborative Innovation Center of Advanced Microstructures. The microfabrication center of the National Laboratory of Solid State Microstructures (NLSSM) is also acknowledged for their technical support.

## Data Availability Statement

The data that support the findings of this study are available from the corresponding author upon reasonable request.

## Conflict of Interest

The authors declare no conflict of interest.

**Table of contents**

ToC text:

This manuscript for the first time proposes and experimentally demonstrates the concept of skin-inspired in-sensor encoding, by exploiting the strain-tunable quantum geometry of electronic bands in van der Waals topological semimetal. This scheme enables a single device to perform high-dimensional sensing and in-sensor encoding of strain vectors, representing a significant advancement in straintronics and biomimetic electronics for quantum materials.

Zenglin Liu, Jingwen Shi, Jin Cao, Zecheng Ma, Zaizheng Yang, Yanwei Cui, Lizheng Wang, Yudi Dai, Moyu Chen, Pengfei Wang, Yongqin Xie, Fanqiang Chen, Youguo Shi, Cong Xiao, Shengyuan A. Yang, Bin Cheng*, Shi-Jun Liang*, Feng Miao*

**Skin-inspired in-sensor encoding of strain vector using tunable quantum geometry**

ToC figure:

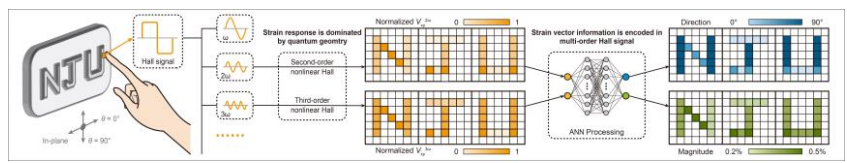

**Skin-inspired in-sensor encoding of strain vector using tunable quantum geometry**

*Zenglin Liu, Jingwen Shi, Jin Cao, Zecheng Ma, Zaizheng Yang, Yanwei Cui, Lizheng Wang, Yudi Dai, Moyu Chen, Pengfei Wang, Yongqin Xie, Fanqiang Chen, Youguo Shi, Cong Xiao, Shengyuan A. Yang, Bin Cheng\*, Shi-Jun Liang\*, Feng Miao\**

1. Calculation results of the two-dimensional tilted Dirac model

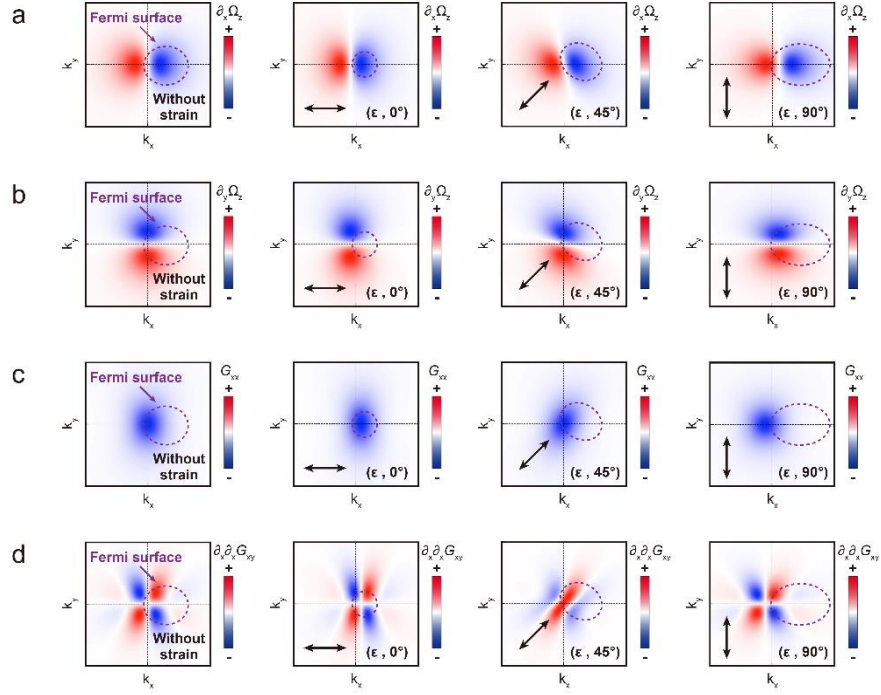

**Figure S1.** The band geometric quantities are calculated from the two-dimensional tilted Dirac model, with the purple dashed line indicating the position of the Fermi surface. a) The distribution of the partial derivative of Berry curvature $\partial_x \Omega_z$ without strain is shown on the left, while the three images on the right depict the distributions of the partial derivative of Berry curvature $\partial_x \Omega_z$ when uniaxial strain vectors (black arrow) are applied at 0°, 45° and 90°. b) The distribution of the partial derivative of Berry curvature $\partial_y \Omega_z$ without strain is shown on the left, while the three images on the right depict the distribution of the partial derivative of Berry curvature $\partial_y \Omega_z$ when uniaxial strain vectors (black arrow) are applied at 0°, 45° and 90°. c) The distribution of the Berry-connection polarizability tensor components $G_{xx}$ without strain is shown on the left, while the three images on the right depict the distributions of the Berry-connection polarizability tensor components $G_{xx}$ when uniaxial strain vectors (black arrow) are applied at 0°, 45° and 90°. d) The distribution of the second partial derivative of Berry-connection polarizability tensor components $\partial_x \partial_x G_{xx}$ without strain is shown on the left, while the three images on the right depict the distribution of the second partial derivative of Berry-connection polarizability tensor components $\partial_x \partial_x G_{xx}$ when uniaxial strain vectors (black arrow) are applied at 0°, 45° and 90°.

2. Sample thickness of Device 01 determined by atomic force microscope

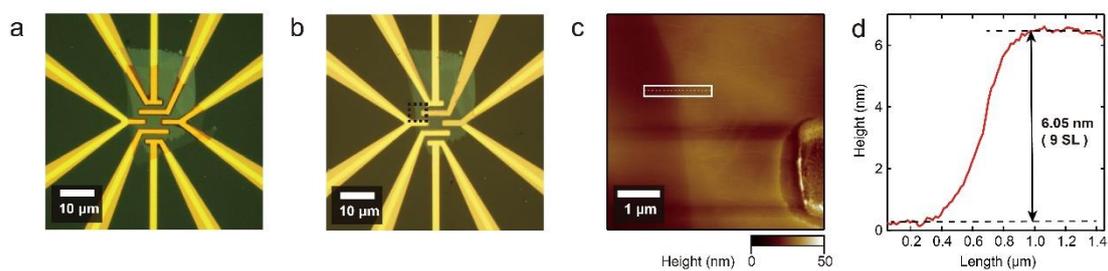

**Figure S2.** a) Optical image of Device 01 after transport measurement. b) Optical image of the exposed Device 01. The thick-layer h-BN, used for surface protection, was removed from the device surface using the method of sample transfer. c) Height profile of the black-boxed region in (b) determined by atomic force microscope. d) The thickness of the WTe$_2$ sample is approximately 6.05 nm, corresponding to 9 layers. Scanning position is represented by a white dashed line in (c).

3. Crystal orientation of the sample determined by polarized Raman spectroscopy

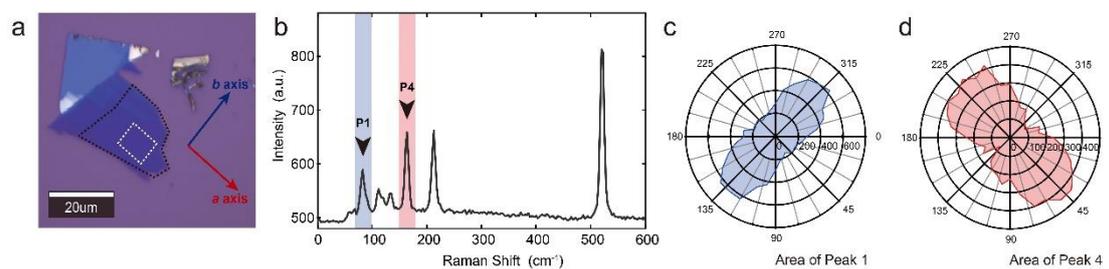

**Figure S3.** a) The optical image of the sample for Device 01 before being transferred to the polyimide substrate. b) Raman spectrum of the WTe$_2$ sample. The areas of Raman peak 1 (P1) and Raman peak 4 (P4) are marked with red and blue shading, respectively. c) The angle-dependent of P1 area. The direction corresponding to the maximum value of the peak area represents the *b*-axis of the sample. d) The angle-dependent of P4 area. The direction corresponding to the maximum value of the peak area represents the *a*-axis of the sample.

4. Homemade strain device used to apply strain vectors

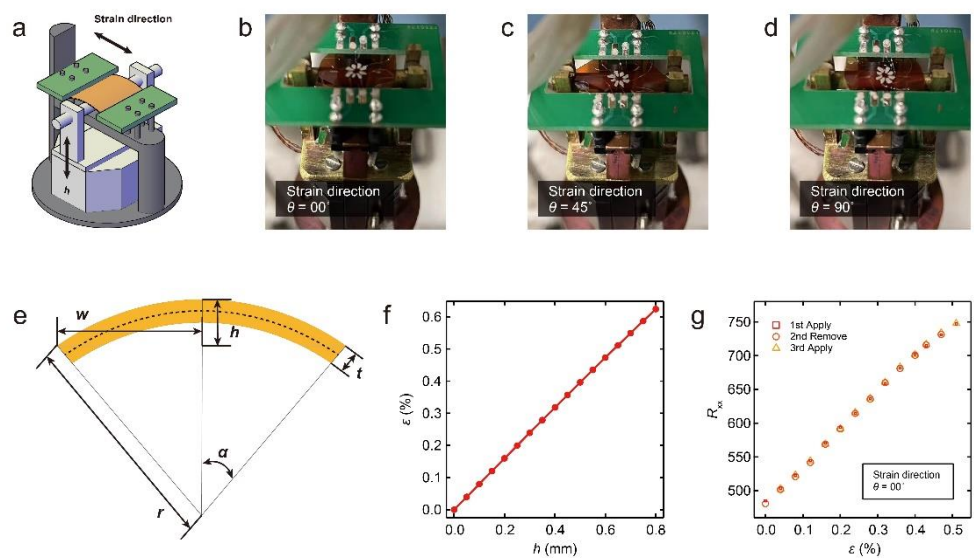

**Figure S4.** a) A structural diagram of the homemade strain setup. b,c,d) Optical photographs depicting the configuration of the homemade strain setup and the device when applying 0° (b), 45° (c), and 90° (d) strain vectors. e) Schematic diagram of polyimide substrate bending. $t$ is the thickness of the polyimide substrate, $r$ is the radius of curvature, $h$ is the lifted height and $w$ is half the distance between two fixed ends (the green block in (a)). f) The relationship between the lifted height and the magnitude of the strain vector. g) The four-probe resistance as a function of the magnitude of strain vector. The repeatability of the resistance changes upon multiple cycles of applying and removing strain demonstrates the robustness of the strain application capability of the homemade strain setup.

## 5. Strain response of harmonic signals

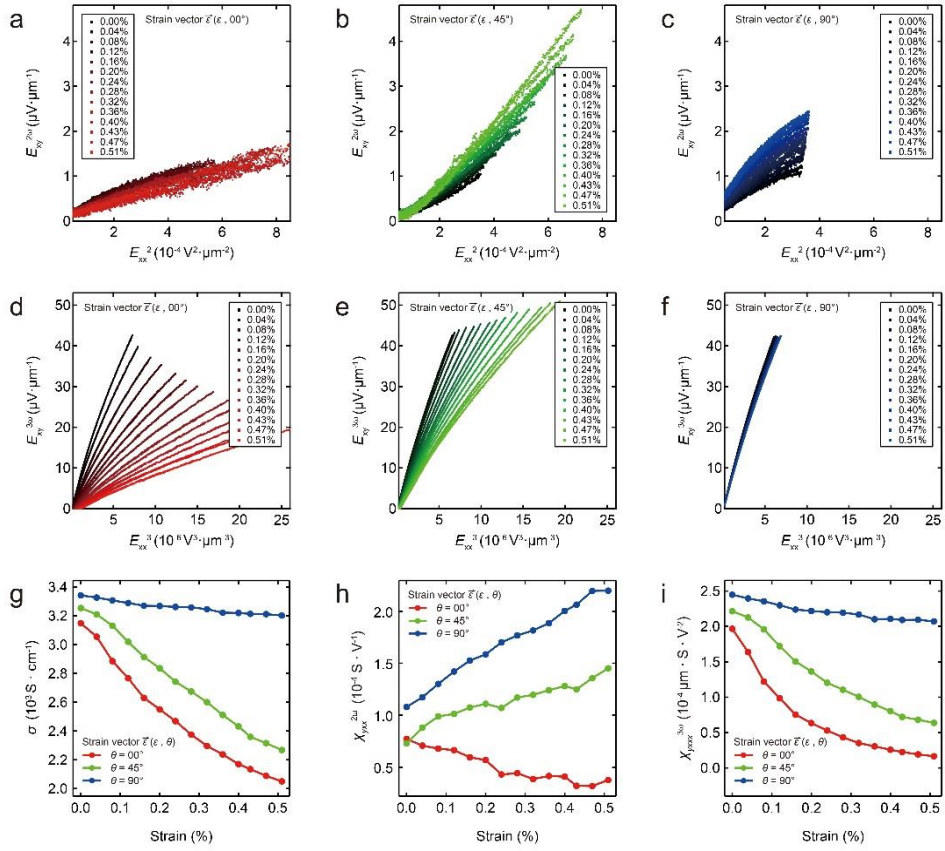

**Figure S5.** a,b,c) Second-harmonic transverse electric field $E_{xy}^{2\omega} = V_{xy}^{2\omega}/W$ is plotted as a function of the square of longitudinal electric field $E_{xx} = V_{xx}^{\omega}/L$ when different magnitudes of strain vectors are applied at 0° (a), 45° (b) and 90° (c). d,e,f) Third-harmonic transverse electric field $E_{xy}^{3\omega} = V_{xy}^{3\omega}/W$ is plotted as a function of the cube of longitudinal electric field $E_{xx}$ when different magnitudes of strain vectors are applied at 0° (d), 45°(e) and 90°(f). g) The longitudinal conductivity $\sigma = I(L/Wd)/V_{xx}^{\omega}$ is plotted as functions of the magnitude of strain vector applied in different directions. h) The second-order nonlinear Hall conductivity $\chi_{yxx}^{2\omega} = J_{yxx}^{2\omega}/(E_{xx}^{\omega})^2$ is plotted as functions of the magnitude of strain vector applied in different directions. i) The third-order nonlinear Hall conductivity $\chi_{yxxx}^{3\omega} = J_{yxxx}^{3\omega}/(E_{xx}^{\omega})^3$ is plotted as functions of the magnitude of strain vector applied in different directions.

6. Strain response measurements performed on other devices

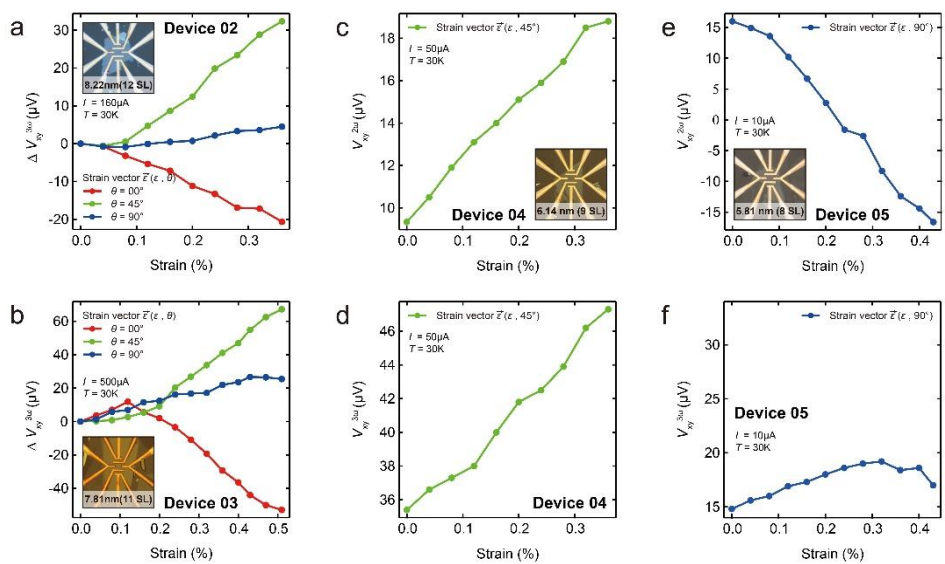

**Figure S6.** a) The variation of the third-harmonic voltage signal from Device 02 is plotted as a function of the magnitude of the strain vector applied in different directions. b) The variation of the third-harmonic voltage signal from Device 03 is plotted as a function of the magnitude of the strain vector applied in different directions. c,d) The second-harmonic (c) and the third-harmonic (d) voltage signals from Device 04 are plotted as a function of the magnitude of the strain vector applied at 45°. e,f) The second-harmonic (e) and the third-harmonic (f) voltage signals from Device 05 are plotted as a function of the magnitude of the strain vector applied at 90°. The insets in (a), (b), (c), and (e) present optical photographs of Devices 02, 03, 04, and 05, respectively.

7. Temperature dependence of harmonic signals

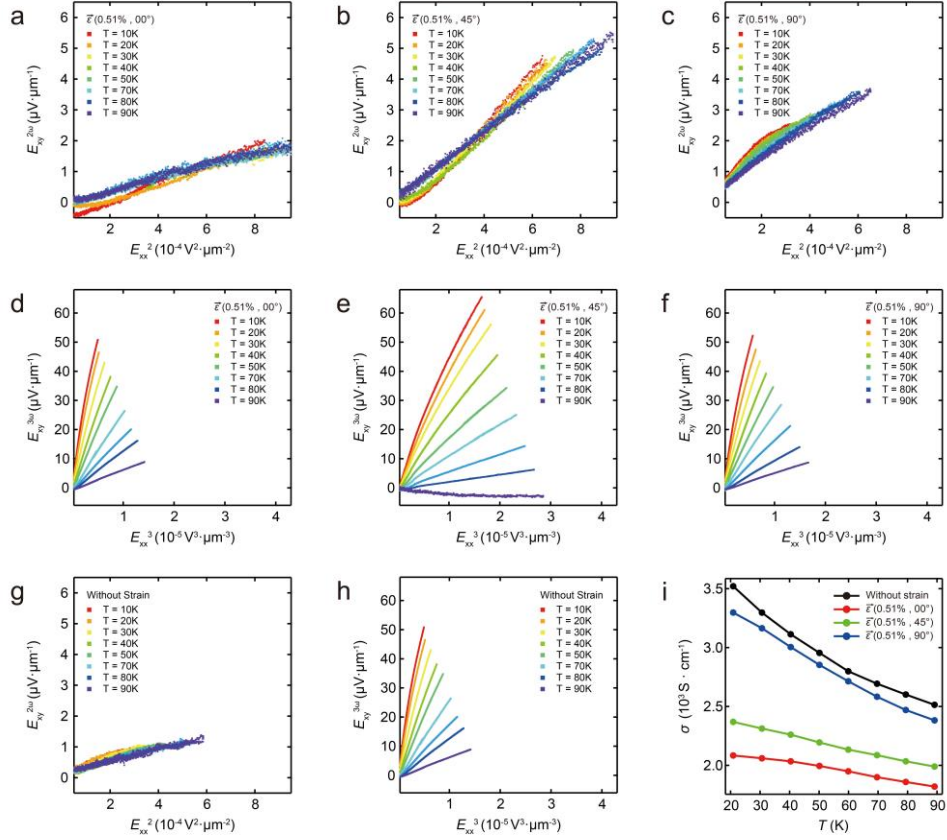

**Figure S7.** a,b,c) Second-harmonic transverse electric field $E_{xy}^{2\omega} = V_{xy}^{2\omega}/W$ is plotted as a function of the square of longitudinal electric field $E_{xx} = V_{xx}^{\omega}/L$ at different temperatures when the 0.51% strain vectors are applied at 0° (a), 45° (b) and 90° (c). d,e,f) Third-harmonic transverse electric field $E_{xy}^{3\omega} = V_{xy}^{3\omega}/W$ is plotted as a function of the cube of longitudinal electric field $E_{xx}$ at different temperatures when the 0.51% strain vectors are applied at 0° (d), 45° (e) and 90° (f). g) Second-harmonic transverse electric field $E_{xy}^{2\omega}$ is plotted as a function of the square of longitudinal electric field $E_{xx}$ at different temperatures when no strain is applied. h) Third-harmonic transverse electric field $E_{xy}^{3\omega}$ is plotted as a function of the cube of longitudinal electric field $E_{xx}$ at different temperatures when no strain is applied. i) The longitudinal conductivity $\sigma = I(L/Wd)/V_{xx}^{\omega}$ is plotted as functions of temperature with strain vectors applied in different directions or without strain.

8. Supplementary measurement data for Device 01

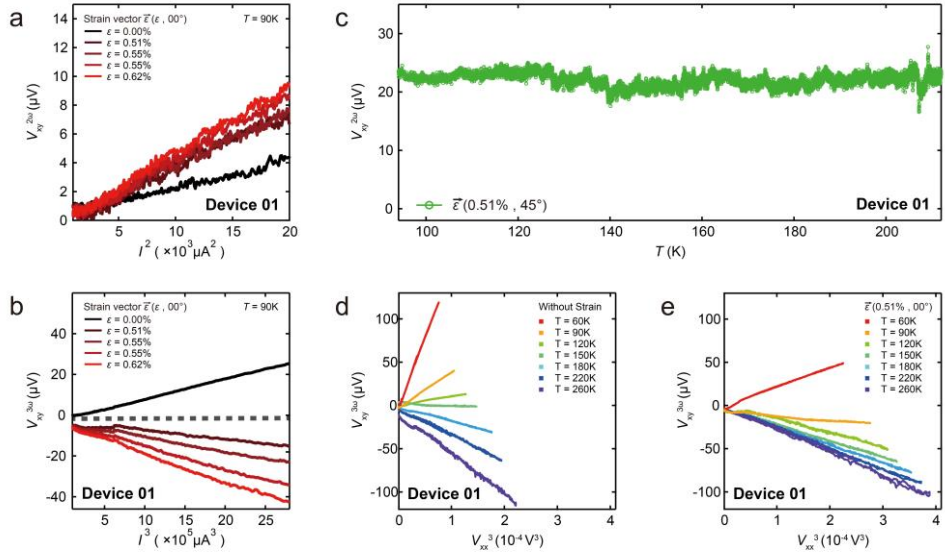

**Figure S8.** a) Second-harmonic transverse voltage signal $V_{xy}^{2\omega}$ depends linearly on the square of AC current amplitude $I$ at 90 K when different magnitudes of strain vectors are applied at 0°. b) Third-harmonic transverse voltage signal $V_{xy}^{3\omega}$ depends linearly on the cube of AC current amplitude $I$ at 90 K when different magnitudes of strain vectors are applied at 0°. c) Second-harmonic transverse voltage signal $V_{xy}^{2\omega}$ is plotted as a function of temperature with the 0.51% strain vector applied at 45°. d,e) Third-harmonic transverse voltage signal $V_{xy}^{3\omega}$ is plotted as a function of temperature without strain (d) or with the 0.51% strain vector applied at 0° (e).

9. Harmonic voltage signals in response to strain vectors

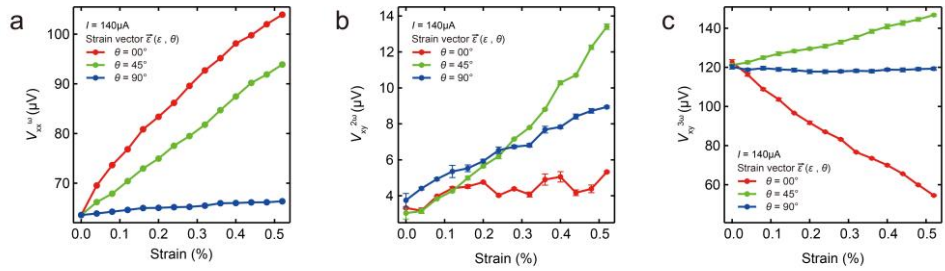

**Figure S9.** a) The longitudinal voltage signal $V_{xx}^{\omega}$ is plotted as functions of the magnitude of strain vector applied in different directions. b) Second-harmonic transverse voltage signal $V_{xy}^{2\omega}$ is plotted as functions of the magnitude of strain vector applied in different directions. c) Third-harmonic transverse voltage signal $V_{xy}^{3\omega}$ is plotted as functions of the magnitude of strain vector applied in different directions. The driving currents are fixed at 140 µA. The error bars are calculated based on the standard deviation associated with the statistical averages of forward and backward scans.

10. Extracting accuracy of high-dimensional information over the epoch

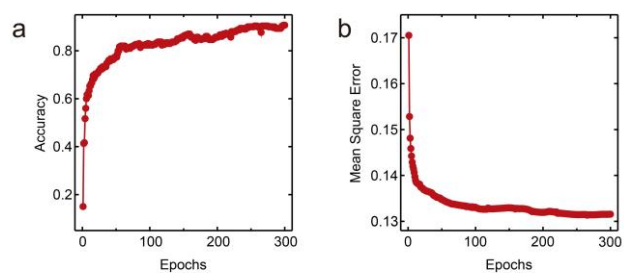

**Figure S10.** a) The accuracy surpasses the 90% with an increment in the number of epochs.
b) The mean square error decreases below 0.14 with an increment in the number of epochs.

11. The encoding of embossed characters

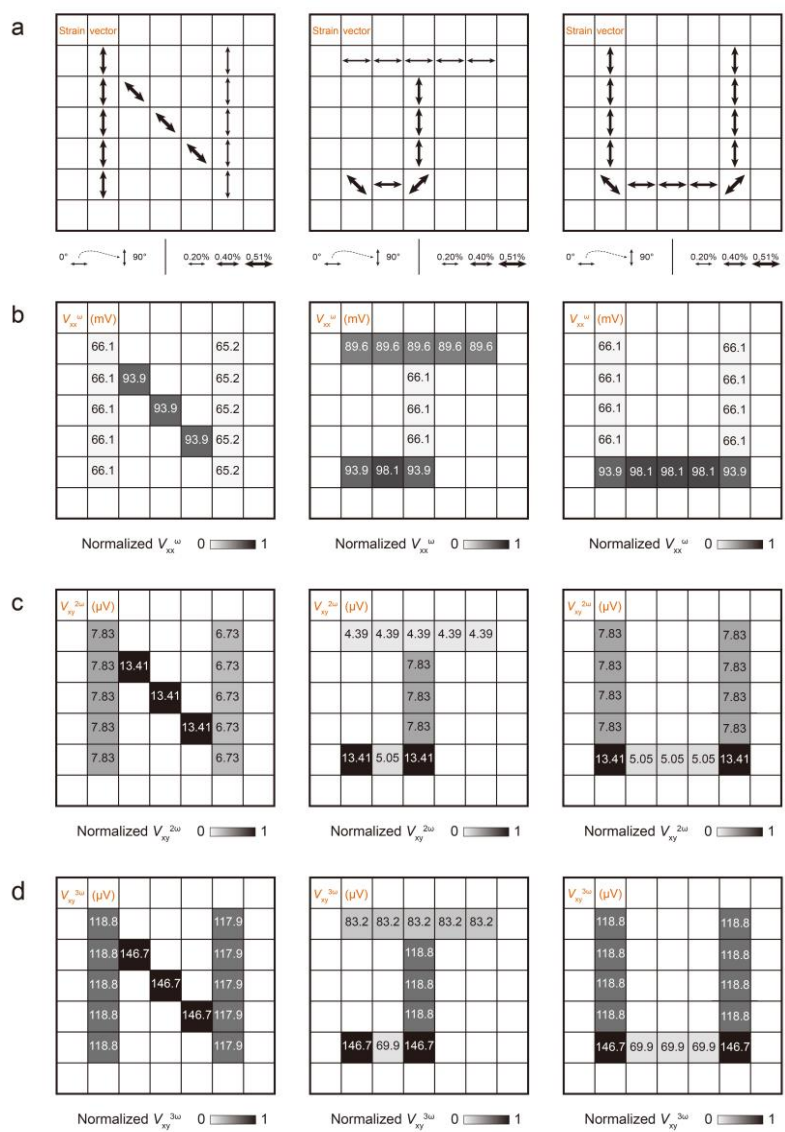

**Figure S11.** Each character in "NJU" is composed of 25 (5×5) pixels. a) Each pixel corresponds to a specific strain vector. b) Each pixel corresponds to the longitudinal voltage signal associated with the piezoresistive effect. c,d) Each pixel corresponds to the second-harmonic (c) and third-harmonic (d) transverse voltage signals associated with the nonlinear Hall effect.

12. Coefficients extracted from the scaling law

**Table 1.**

| Quantity and symbol | Strain vector | Value with unit |
|---|---|---|
| The slope derived from the fitted $E_{xy}^{2\omega}/E_{xx}^{2}$ versus $\sigma^2$ curve: $\xi^{2\omega}$ | None | $1.951 \times 10^{-8}\ \mu m^3 \cdot V^{-1} \cdot S^{-2}$ |
| | (0.51%, 0°) | $9.772 \times 10^{-9}\ \mu m^3 \cdot V^{-1} \cdot S^{-2}$ |
| | (0.51%, 45°) | $6.214 \times 10^{-8}\ \mu m^3 \cdot V^{-1} \cdot S^{-2}$ |
| | (0.51%, 90°) | $4.094 \times 10^{-8}\ \mu m^3 \cdot V^{-1} \cdot S^{-2}$ |
| The intercept derived from the fitted $E_{xy}^{2\omega}/E_{xx}^{2}$ versus $\sigma^2$ curve: $\eta^{2\omega}$ | None | $6.554 \times 10^{-4}\ \mu m \cdot V^{-1}$ |
| | (0.51%, 0°) | $1.592 \times 10^{-3}\ \mu m \cdot V^{-1}$ |
| | (0.51%, 45°) | $3.190 \times 10^{-3}\ \mu m \cdot V^{-1}$ |
| | (0.51%, 90°) | $3.166 \times 10^{-3}\ \mu m \cdot V^{-1}$ |
| The fitted slope derived from the $E_{xy}^{3\omega}/E_{xx}^{3}$ versus $\sigma^2$ curve: $\xi^{3\omega}$ | None | $1.454 \times 10^{-4}\ \mu m^4 \cdot V^{-2} \cdot S^{-2}$ |
| | (0.51%, 0°) | $1.012 \times 10^{-4}\ \mu m^4 \cdot V^{-2} \cdot S^{-2}$ |
| | (0.51%, 45°) | $2.434 \times 10^{-4}\ \mu m^4 \cdot V^{-2} \cdot S^{-2}$ |
| | (0.51%, 90°) | $1.417 \times 10^{-4}\ \mu m^4 \cdot V^{-2} \cdot S^{-2}$ |
| The intercept derived from the fitted $E_{xy}^{3\omega}/E_{xx}^{3}$ versus $\sigma^2$ curve: $\eta^{3\omega}$ | None | $-8.653\ \mu m^2 \cdot V^{-2}$ |
| | (0.51%, 0°) | $-3.462\ \mu m^2 \cdot V^{-2}$ |
| | (0.51%, 45°) | $-9.879\ \mu m^2 \cdot V^{-2}$ |
| | (0.51%, 90°) | $-7.701\ \mu m^2 \cdot V^{-2}$ |